\documentclass[aps,prl,reprint,groupedaddress,showpacs,floatfix]{revtex4-1}

\usepackage{graphicx}
\usepackage{dcolumn}
\usepackage{bm}
\usepackage[mathlines]{lineno}

\begin{document}

\title{\hfill {\small\rm Phys. Rev. Lett. {\bf 112}, 026803 (2014)}\\
       Topologically protected conduction state at
       carbon foam surfaces: \\
       An \textit{ab-initio} study}

\author{Zhen Zhu}
\affiliation{Physics and Astronomy Department,
             Michigan State University,
             East Lansing, Michigan 48824, USA}

\author{Zacharias G. Fthenakis}
\affiliation{Physics and Astronomy Department,
             Michigan State University,
             East Lansing, Michigan 48824, USA}

\author{Jie Guan}
\affiliation{Physics and Astronomy Department,
             Michigan State University,
             East Lansing, Michigan 48824, USA}

\author{David Tom\'{a}nek}
\email[E-mail: ]{tomanek@pa.msu.edu}%
\affiliation{Physics and Astronomy Department,
             Michigan State University,
             East Lansing, Michigan 48824, USA}

\date{\today} 

\begin{abstract}
We report results of \textit{ab initio} electronic structure and
quantum conductance calculations indicating the emergence of
conduction at the surface of semiconducting carbon foams.
Occurrence of new conduction states is intimately linked to the
topology of the surface and not limited to foams of elemental
carbon. Our interpretation based on rehybridization theory
indicates that conduction in the foam derives from first- and
second-neighbor interactions between $p_\|$ orbitals lying in the
surface plane, which are related to $p_\perp$ orbitals of
graphene. The topologically protected conducting state occurs on
bare and hydrogen-terminated foam surfaces and is thus unrelated
to dangling bonds. Our results for carbon foam indicate that the
conductance behavior may be further significantly modified by
surface patterning.
\end{abstract}



\maketitle

There is growing interest in topologically complex carbon foam
structures%
\cite{{ANIEfoam13},{Zhu-foam2011},{Ajayan12},{Chenthree2011},%
{DT141},{Seifert2006},{Jiang13},{DT215}} that contain both $sp^2$
and $sp^3$ hybridized atoms. Most experimental studies have
focussed on the synthesis and structural characterization of
carbon foams\cite{{Ajayan12},{Zhu-foam2011},{Chenthree2011}} that
are closely related to previously postulated
structures\cite{{DT141},{Seifert2006},{DT215}}. Electronic
properties of
foams\cite{{ANIEfoam13},{DT141},{Seifert2006},{Jiang13}} have
received much less attention than their structural stability in
spite of the obvious possibility to fine-tune the fundamental band
gap value in-between zero in $sp^2-$bonded graphene
and $5.5$~eV in $sp^3-$bonded diamond by modifying the foam
morphology.

Here we report results of \textit{ab initio} electronic structure
and quantum conductance calculations indicating the emergence of
conduction at the surface of semiconducting carbon foams.
Occurrence of new conduction states in these systems is intimately
linked to the topology of the surface and not limited to foams of
elemental carbon. Our interpretation based on rehybridization
theory indicates that conduction in the foam derives from first-
and second-neighbor interactions between $p_\|$ orbitals lying in
the surface plane, which are related to $p_\perp$ orbitals of
graphene. The topologically protected conducting state occurs on
bare and hydrogen-terminated foam surfaces and is thus unrelated
to dangling bonds. Our results for carbon foam indicate that the
conductance behavior may be further significantly modified by
surface patterning.

\begin{figure}[tb]
\includegraphics[width=1.0\columnwidth]{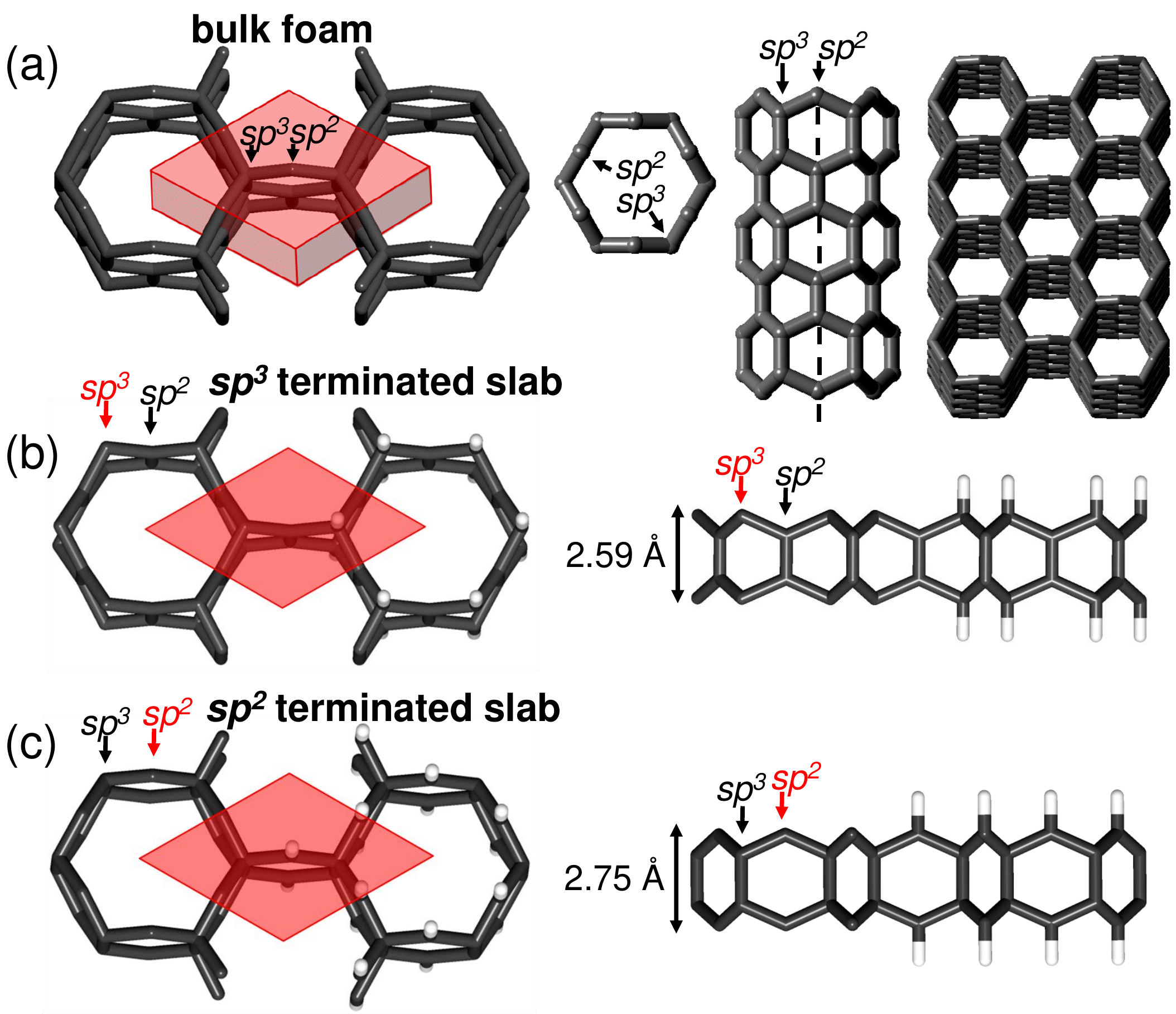}
\caption{(Color online) Geometry of carbon foam and thin foam
slabs. (a) Structure of bulk foam and that of an individual foam
cell in side and top view, allowing to distinguish $sp^2$ and
$sp^3$ sites. The dashed line shows the long cell axis. The
perspective view in the right panel depicts a larger bulk segment.
Structure of (b) an $sp^3-$terminated and (c) an $sp^2-$terminated
thin foam slab. The tilted view used in left panels depicts the
structure and primitive unit cells. The right panels in (b) and
(c) are side views of the structure that better illustrate the
type of termination and illustrate partial hydrogen coverage.
\label{fig1}}
\end{figure}

The bulk carbon foam, depicted in Fig.~\ref{fig1}(a), is a
cellular structure resembling vaguely a fused triangular array of
(6,0) zigzag nanotubes. In contrast to a nanotube array, the walls
of the foam cells consist of 60\% $sp^2$ bonded atoms shared by
two neighboring cells and 40\% $sp^3$ bonded atoms shared by three
adjacent cells. Density-functional based tight-binding (DFTB)
results indicate that the bulk structure is a
semiconductor\cite{Seifert2006} with a band gap of 2.55~eV.

Cleavage normal to the long cell axis may generate two different
surfaces. We distinguish the $sp^3$ surface terminated by C atoms,
which were fourfold coordinated in the bulk, from the $sp^2$
surface terminated by atoms that were threefold coordinated. For
computational reasons, we will represent the surface by slabs of
finite thickness with two identical surfaces that are either bare
or terminated by hydrogen.
The optimum structure of the thinnest foam slab with $sp^3$
termination is shown in Fig.~\ref{fig1}(b) and that of the
thinnest $sp^2-$terminated slab in Fig.~\ref{fig1}(c). We will
focus on these thinnest slabs in the main manuscript and
demonstrate the generality of our results for thicker slabs in the
Supplemental Material~\cite{SM-fmcond13}.

Our numerical results for the equilibrium structure, stability and
electronic properties of carbon foam slabs are based on density
functional theory (DFT) as implemented in the \textsc{SIESTA}
code~\cite{SIESTA}. The different foam surfaces are represented by
a periodic array of slabs, separated by a 15~{\AA} thick vacuum
region. We used the Ceperley-Alder~\cite{Ceperley1980}
exchange-correlation functional as parameterized by Perdew and
Zunger~\cite{Perdew81}, norm-conserving Troullier-Martins
pseudopotentials~\cite{Troullier91}, and a double-$\zeta$ basis
including polarization orbitals. The reciprocal space was sampled
by a fine grid~\cite{Monkhorst-Pack76} of
$16{\times}16{\times}1$~$k$-points in the Brillouin zone of the
primitive surface unit cell. We used a mesh cutoff energy of
$180$~Ry to determine the self-consistent charge density, which
provided us with a precision in total energy of
${\alt}2$~meV/atom.

Transport properties of the slabs were investigated using the
nonequilibrium Green's function (NEGF) approach as implemented in
the \textsc{TRAN-SIESTA} code \cite{transiesta}. For structures
optimized by DFT, we used a single-$\zeta$ basis with polarization
orbitals, the same $180$~Ry mesh cutoff energy, and a
$8{\times}8{\times}1$~$k$-point grid~\cite{Monkhorst-Pack76}.

\begin{figure}[tb]
\includegraphics[width=1.0\columnwidth]{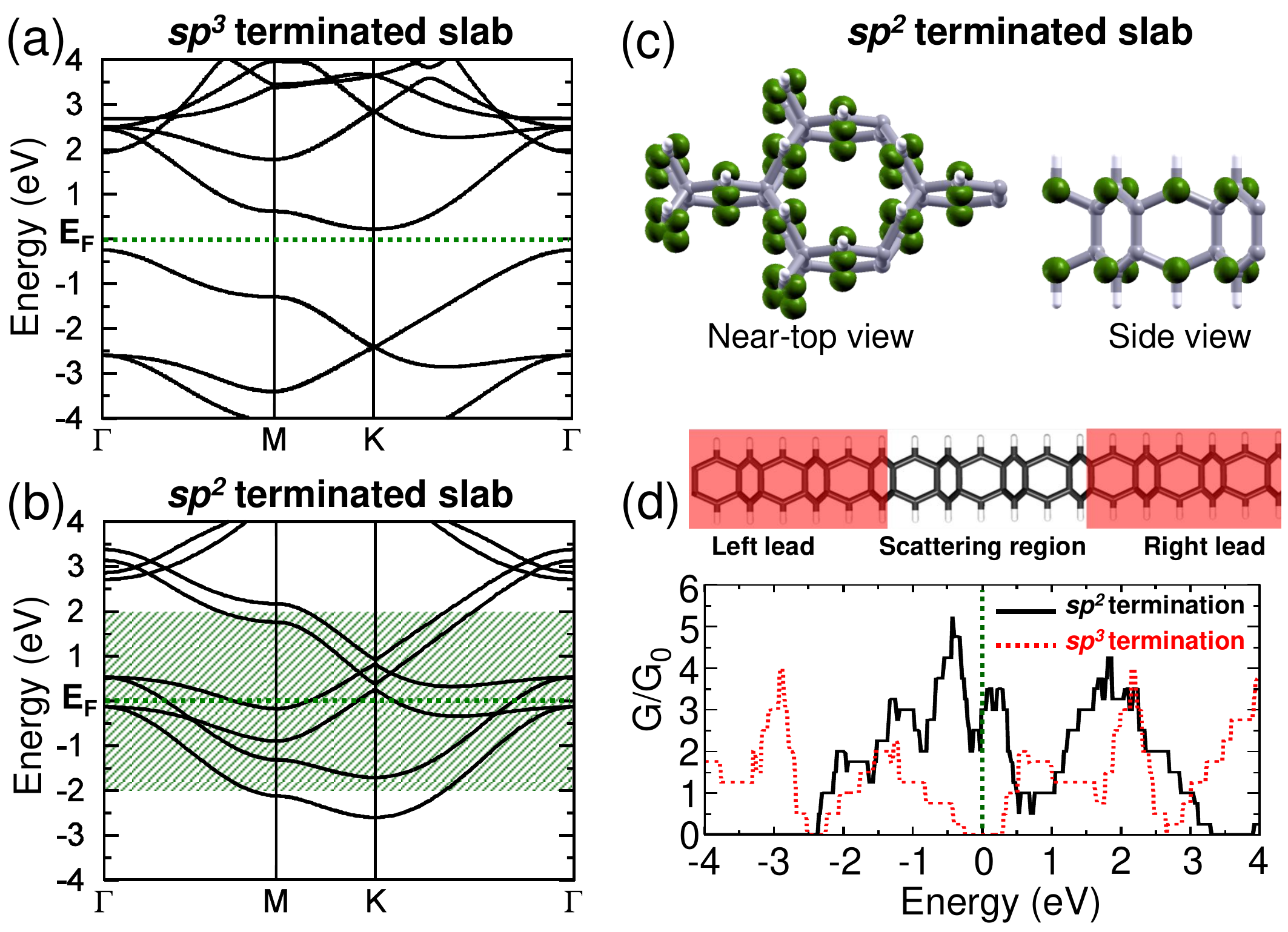}
\caption{(Color online) Electronic structure of thin foam slabs.
DFT-based band structure of a thin, hydrogen-covered foam slab
with (a) $sp^3$ and (b) $sp^2$ termination on both sides. (c)
Charge distribution in the $sp^2-$terminated slab corresponding to
states in the energy range $E_F-2$~eV$<E<E_F+2$~eV, indicated by
shading in (b). (d) Conductance $G$ of $sp^3-$ and
$sp^2-$terminated slabs along the armchair direction, in units of
the conduction quantum $G_0$. \label{fig2}}
\end{figure}

Our DFT results for the band structure of the hydrogen-covered
thin carbon slab with two $sp^3$ surfaces, shown in
Fig.~\ref{fig1}(b), are presented in Fig.~\ref{fig2}(a). These
results suggest this system to be
semiconducting\cite{DFT-band-gaps}, same as its bulk
counterpart\cite{Seifert2006}. On the other hand, the H-covered,
$sp^2-$terminated slab, depicted in Fig.~\ref{fig1}(c), is clearly
metallic according to the band structure results in
Fig.~\ref{fig2}(b). This result is surprising, since conduction in
semiconducting carbon structures including diamond has so far only
been observed in presence of unsaturated dangling
bonds\cite{OkadaJPSJ13}. Our band structure results in
Fig.~\ref{fig2}(b) also show a Dirac-like cone similar to graphene
at the $K-$point in the Brillouin zone. Moderate $n-$doping should
be able to align it with the Fermi level, providing carriers with
the same desirable properties as graphene.

To learn more about the character of the new conduction states, we
display in Fig.~\ref{fig2}(c) the charge density associated with
states close to $E_F$, shown by the shaded region in
Fig.~\ref{fig2}(b). These states with $p_\|$ character, which are
oriented within the surface plane of the foam, are located only on
the $sp^2$ sublattice. This is very different from graphene, where
conduction is caused by nearest-neighbor hopping between $p_\perp$
orbitals oriented normal to the surface, which are equally
occupied at all lattice sites.

To judge the suitability of carbon foam for electronic
applications, we calculated quantum conductance of
hydrogen-covered thin $sp^2-$ and $sp^3-$terminated carbon foam
slabs, shown in Fig.~\ref{fig1}(b-c), and present the results in
Fig.~\ref{fig2}(d). These results reflect our band structure
results, namely a large conductance at small bias values in the
$sp^2-$terminated carbon foam slab and a conduction gap of
$0.6$~eV in the $sp^3-$terminated slab. Additional
results\cite{SM-fmcond13}, also for thicker slabs,
suggest that conduction is linked to the $sp^2-$termination of the
carbon foam surface and nearly isotropic.

\begin{figure*}[tb]
\includegraphics[width=2.0\columnwidth]{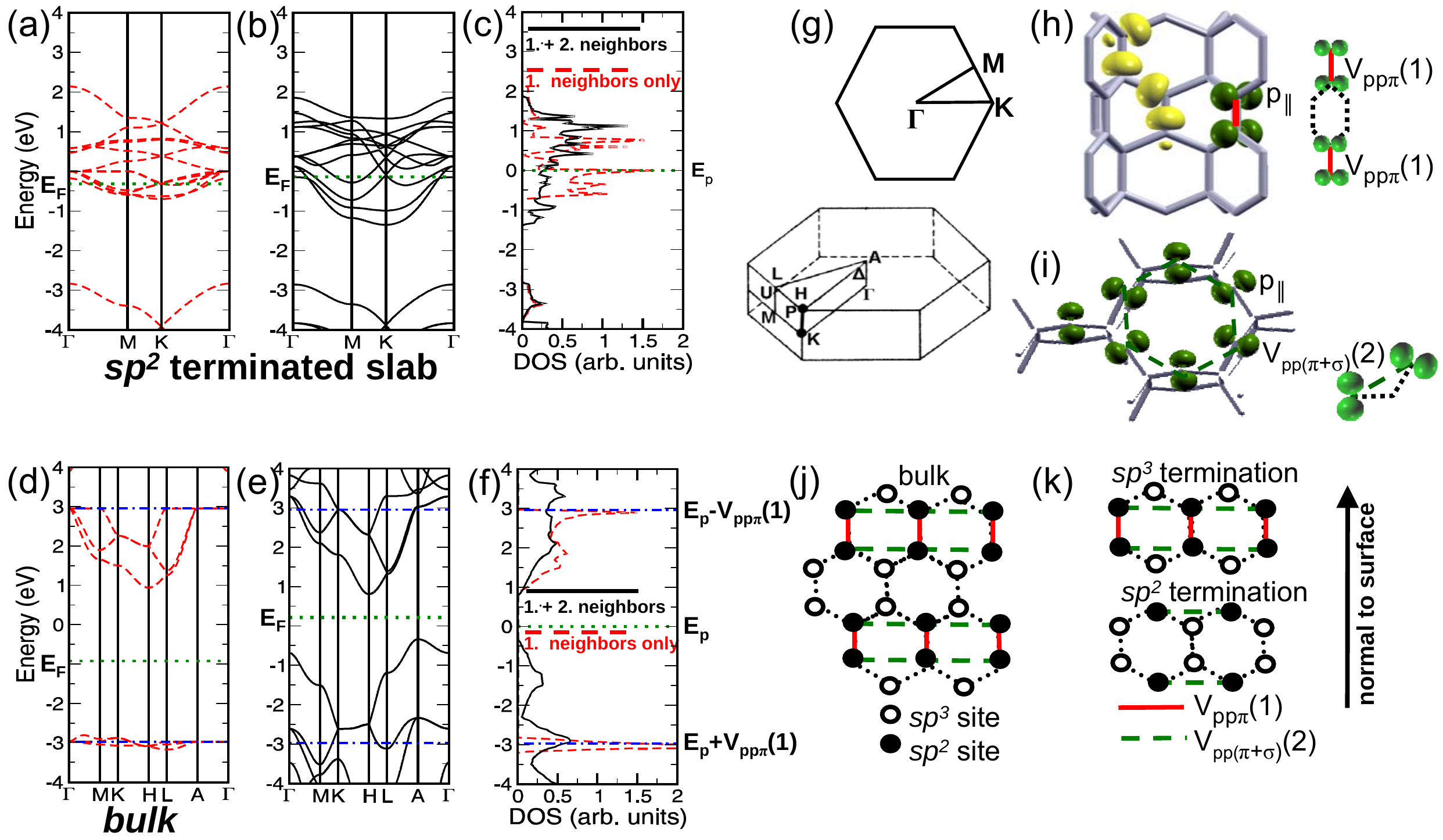}
\caption{(Color online) Interpretation of electronic structure
results for the $sp^2-$terminated slab (a-c) and bulk foam (d-f)
using the LCAO technique. Band structure results considering only
nearest-neighbor interaction (a,d), results of calculations that
also include second-neighbor interaction between $p_\|$ states
(b,e), and the corresponding densities of states (c,f). (g)
High-symmetry points in the slab and bulk Brillouin zones. (h)
Rehybridized orbitals used in the calculation. $p_\|$ orbitals,
shown in darker (green) shade, are responsible for conduction. (i)
Tilted top view of the second neighbor-interaction between $p_\|$
states that dominate band dispersion near $E_F$. Connectivity
diagram for the bulk (j) and thin foam slabs (k) that helps
explain the semiconducting behavior of bulk and $sp^3-$terminated
foam, and the origin of the conducting state at the
$sp^2-$terminated surface. \label{fig3}}
\end{figure*}

To confirm the generality of this finding and obtain insight into
its origin, we used the linear combination of atomic orbitals
(LCAO) technique\cite{{Slater-Koster54},{DT049},{Papacon}} to
study the electronic structure of carbon foam
slabs\cite{SM-fmcond13}. Our results for the $sp^2$ terminated
slab of interest are presented in
Figs.~\ref{fig3}(a)-\ref{fig3}(c). The band associated with
conduction, obtained by considering only nearest-neighbor
interactions and presented in Fig.~\ref{fig3}(a), was found to be
quite different from its DFT-based counterpart in
Fig.~\ref{fig2}(b). To find out if this difference is caused by
omitting second and third neighbor interactions, we introduced
these interactions in our initial Hamiltonian\cite{SM-fmcond13}.
By selectively modifying individual hopping
parameters, we have found that %
(i) the most significant changes in the electronic structure,
including band broadening near $E_F$, are caused by
second-neighbor $V_{pp\sigma}(2)$ and $V_{pp\pi}(2)$ interactions; %
(ii) $V_{ss\sigma}(2)$ and $V_{sp\sigma}(2)$ interactions between
second neighbors do not affect the electronic structure near $E_F$
and can be neglected; %
(iii) third-neighbor interactions contribute very little to the
electronic structure and can also be safely neglected. %
Results obtained using the Hamiltonian augmented by
second-neighbor interactions between $p_\|$
states\cite{SM-fmcond13}, presented in Fig.~\ref{fig3}(b), agree
much better with the DFT results of Fig.~\ref{fig2}(b), in
particular regarding the width of the occupied part of the
conduction band.

The effect of the second-neighbor interaction is even more
pronounced in the electronic structure of the bulk foam, shown in
Figs.~\ref{fig3}(d)-\ref{fig3}(f). As seen from the comparison
between Figs.~\ref{fig3}(d) and \ref{fig3}(e), neglecting the
second-neighbor interaction in Fig.~\ref{fig3}(d) caused a drastic
narrowing of the bulk valence band and a significant increase in
the fundamental band gap. LCAO calculations for the bulk with both
first and second neighbor interactions, presented in
Fig.~\ref{fig3}(e), indicate an indirect $A-H$ fundamental band
gap of $E_g=1.12$~eV, in qualitative agreement with
non-self-consistent DFTB results.

Having shown that the electronic structure near the Fermi level is
well reproduced by the LCAO Hamiltonian, which also considers
second-neighbor interactions between $p$ states, we proceed to
identify the reason for the fundamental difference between the
$sp^2-$ terminated metallic and $sp^3-$ terminated semiconducting
slab. We use the rehybridization theory\cite{{Haddon1},{Haddon2}}
to identify proper hybrid orbitals, since the atomic arrangements
in the foam are generally not purely linear, hexagonal or
tetrahedral. Hybrid orbitals $|h_i\rangle$ with $i=1,2,3,4$, which
are associated with an atom, are linear combinations of the
$|s\rangle$ and a $|p\rangle$ atomic orbital at this site. The
direction of the $|p\rangle$ orbital is not that of the Cartesian
axes, but rather taken as that of nearest-neighbor bonds to up to
three neighbors. The relative weights of the atomic orbitals are
constructed\cite{Fthenakis} by enforcing the orthogonality
condition ${\langle}h_i|h_j{\rangle}=\delta_{ij}$. This
orthogonality condition is also used to construct any remaining
hybrid orbitals. Selected hybrids at fourfold coordinated $sp^3$
sites and at threefold coordinated $sp^2$ sites are shown in
Fig.~\ref{fig3}(h).

Analysis of the foam eigenstates at $E_F$ indicates dominance of
$p_\|$ hybrids, highlighted by the darker color
Fig.~\ref{fig3}(h), at $sp^2$ sites only. While parallel to the
slab surface, these $p_\|$ orbitals are not aligned along a
spatially fixed direction, but rather are locally normal to the
graphitic strips lining the foam cells. These $p_\|$ hybrids on
the $sp^2$ sublattice play the role of $\pi$ orbitals in the
H\"uckel Hamiltonian that describes most of the interesting
physics in this system. We will show below that considering only
first- and second-neighbor interaction in the H\"uckel Hamiltonian
is sufficient to explain, why bulk foam and $sp^3-$terminated
slabs are semiconducting, whereas $sp^2-$terminated slabs become
metallic.

The spatial distribution of the $p_\|-$dominated states at $E_F$
at one of the slab surfaces, shown in Fig.~\ref{fig3}(i), agrees
well with the DFT results presented in Fig.~\ref{fig2}(c). In a
H\"uckel Hamiltonian with up to second-neighbor interactions, the
$p_\|$ hybrids may interact in two ways only. The first type of
interaction is normal to the surface and involves pairs of
adjacent $sp^2$ sites, which interact by the first-neighbor
$V_{pp\pi}(1)$ interaction. This is shown schematically by the red
solid lines in the right panel of Fig.~\ref{fig3}(h) and in
Figs.~\ref{fig3}(j-k), which represent the connectivity diagram of
the foam. The second type of interaction is in-plane and much
weaker, involving only the second-neighbor $V_{pp\pi}(2)$ and
$V_{pp\sigma}(2)$ interactions between $p_\|$ states. This is
shown schematically by the green dashed lines in the right panel
of Fig.~\ref{fig3}(i) and in Figs.~\ref{fig3}(j-k). In contrast to
graphene, where the relatively strong $V_{pp\pi}(1)$ interaction
forms a network that connects all atoms in the layer, this
interaction connects only pairs of adjacent $sp^2$ sites in the
carbon foam, as illustrated in Figs.~\ref{fig3}(j-k).

We conclude that bulk foam states near $E_F$ resemble those of a
set of carbon dimers, where the $V_{pp\pi}(1)$ interaction splits
the $E_p$ energy eigenvalue into $E_p{\pm}V_{pp\pi}(1)$, indicated
by the blue dash-dotted lines in Figs.~\ref{fig3}(d-f). This
corresponds to opening up a large fundamental band gap with
$E_g=2|V_{pp\pi}(1)|$, which turns carbon foam to a semiconductor.
The $\delta-$function-like localized states at
$E_p{\pm}V_{pp\pi}(1)$, originating from decoupled dimers, are
clearly visible in the density of states of the system with
nearest-neighbor interactions only, shown by the red dashed line
in Fig.~\ref{fig3}(f). Much weaker second-neighbor $V_{pp\pi}(2)$
and $V_{pp\sigma}(2)$ interactions couple these dimers and broaden
the localized states into wide valence and conduction bands, shown
by the solid black lines in the density of states in
Fig.~\ref{fig3}(f).

Now it is rather straight-forward to explain the fundamental
difference between the bulk, $sp^3-$ and $sp^2-$terminated
surfaces in terms of conductivity. Deciding whether a structure is
metallic or semiconducting boils down to the simple question,
whether {\em all} of the $sp^2$ sites are connected as
first-neighbors to another $sp^2$ site. If so, then all energy
eigenvalues $E_p$ of the $p_\|$ states will split by
$V_{pp\pi}(1)$ into the eigenvalue pair $E_p{\pm}V_{pp\pi}(1)$,
which opens up a gap. This is the case for the bulk and an
$sp^3-$terminated surface of the foam. If, on the other hand,
there are at least {\em some} $sp^2$ sites present with no
first-neighbor bonds to other $sp^2$ sites, then the energy
eigenvalue $E_p$ of these $p_\|$ states will not split. Presence
of partly filled $p_\|$ states at $E_p=E_F$ indicates that such a
system should be conducting. The second-neighbor interaction
between $p_\|$ states provides a weak coupling between $sp^2$ site
pairs or lone $sp^2$ sites at the surface, as shown schematically
in Figs.~\ref{fig3}(j) and \ref{fig3}(k). In the bulk or in the
$sp^3-$terminated foam, this second-neighbor interaction broadens
the pair of sharp eigenvalues into a valence and conduction band
that are still separated by a band gap, keeping their
semiconducting character. At the $sp^2-$terminated surface, the
weak second-neighbor interaction between $p_\|$ states at $sp^2$
sites with no $sp^2$ nearest neighbors will broaden the state at
$E_p$ to a metallic band, as seen also in the density of states
presented in Fig.~\ref{fig3}(c). This behavior is not restricted
to carbon foam, but should also occur at surfaces of isomorphic
foams of Si, Si-carbides or BN systems\cite{silicene-foam}.

This reasoning also explains, why conductance depends only on the
presence of $sp^2$ sites with a particular topological
arrangement. We may thus conclude that the conducting state at
$sp^2-$terminated surfaces is topologically protected and is
independent of the fact that the surface has been represented by a
finite-thickness slab. This finding is furthermore confirmed by
our electronic structure and conductance results for a thicker
slab\cite{SM-fmcond13}.
With increasing thickness, the electronic structure of finite
slabs approaches that of the bulk material\cite{DFT-band-gaps}.

Our conclusions regarding occurrence of states at $E_F$ are valid
not only for perfect $sp^2-$terminated surfaces, but also for all
defective structures that contain $sp^2$ sites with no $sp^2$
nearest neighbors. Isolated monatomic vacancies in the bulk will
produce corresponding mid-gap defect states. Lithographic
patterning of the $sp^3-$terminated semiconducting surface by
selective removal of lines of surface atoms will yield lines with
$sp^2-$termination that will behave as conductive nanowires.

In conclusion, we have identified theoretically an unusual
conduction mechanism occurring at the $sp^2-$terminated surface of
a semiconducting carbon foam. To obtain microscopic insight into
the origin of this mechanism, we augmented {\em ab initio}
electronic structure and quantum conductance calculations by
rehybridization theory calculations. We found that the occurrence
of new conduction states in this system is intimately linked to
the topology of the surface and not limited to foams of elemental
carbon. Our interpretation based on rehybridization theory
indicates that conduction in the foam derives from first- and
second-neighbor interactions between $p_\|$ orbitals lying in the
surface plane, which are related to $p_\perp$ orbitals of
graphene. The topologically protected conducting state occurs on
bare and hydrogen-terminated foam surfaces and is thus unrelated
to dangling bonds. Our results for carbon foam indicate that the
conductance behavior may be further significantly modified by
surface patterning, allowing to create conductive paths at the
surface of the semiconducting foam matrix.


\begin{acknowledgements}
This study was supported by the National Science Foundation
Cooperative Agreement \#EEC-0832785, titled ``NSEC: Center for
High-rate Nanomanufacturing''. Computational resources have been
provided by the Michigan State University High Performance
Computing Center.
\end{acknowledgements}


%


\begin{thebibliography}{24}
\bibitem [{sil()}]{silicene-foam}%
  \BibitemOpen
  \href@noop {} {}\bibinfo {note} {Whereas carbon foam is less stable than
  graphene, the foam structure of silicon is more stable than
  silicene.}\BibitemShut {Stop}%
\end{thebibliography}
\end{document}